\documentclass[%
 reprint,
 amsmath,amssymb,
 aps,
]{revtex4-2}
\usepackage{slashed}
\usepackage{graphicx}% Include figure files
\usepackage{dcolumn}% Align table columns on decimal point
\usepackage{bm}% bold math
\begin{document}

\preprint{APS/123-QED}

\title{Anomalous function of a Lorentz-violating QED effective action and the relation between compact bulk scalar propagator and path integral duality}% Force line breaks with \\

\author{Huangcheng Yin}
\altaffiliation{School of Mathematics and Physics, Xi'an Jiaotong-Liverpool University. Suzhou, China \\ Email: Huangcheng.Yin20@student.xjtlu.edu.cn}

\date{\today}% It is always \today, today,

\begin{abstract}
In this paper, we consider a compact five dimensional spacetime with the structure $\mathcal{M}^{1,3}\times S^{1}$. Generally speaking, motion on such a structure will break Lorentz invariance, allowing for causal bulk signals to propagate superluminally. Based on recent articles, we calculate the anomalous function of a gauge invariant but Lorentz-violating term in the $4D$ QED effective action by using path integral. Finally, we find that the compact bulk scalar propagator and path integral duality are consistent, this result brings a new perspective: the behavior of breaking Lorentz invariance caused by dimensional compactness can be seen as path integral duality.
\end{abstract}

\maketitle

%\tableofcontents

\section{\label{sec:level1}Introduction}
The compactification of a single  dimension with a radius of $R$(i.e., $\mathcal{M}^{1,3}\times S^{1}$) is the simplest form of braneworld scenario to consider, Minkowski metric can be induced on braneworld's volume. In the past papers, superluminal propagation has been considered on a
moving braneworld\cite{Greene_2022}\cite{Polychronakos_2023}\cite{bernardo2023brane}, and it's causality\cite{Greene_2023}. Faster than light will cause Lorentz violate, although there is no local indication of the violation, but worldvolume Lorentz invariance is broken globally by the compactification. However, causality remains intact, inherited from the causality of the
underlying $5D$ spacetime. In  \cite{kabat2023induced}, they consider the effective action on the brane induced by loops of bulk fields, a variety of self-energy and vertex corrections due to bulk scalars and gravitons and show that bulk loops with non-zero winding generate UV-finite Lorentz-violating terms in the 4-D effective action.  These terms are part of the Standard Model extension, a general framework for Lorentz-violating effective field theories developed in\cite{Colladay_1997}\cite{Colladay_1998}. There are stringent experimental bounds on the Lorentz-violating coefficients which have been tabulated in \cite{Kosteleck__2011}.\\
By using the fundamental constants $G$, $\hbar$ and $c$, the length with dimensions is $L_{0}=\sqrt{G\hbar/c^{3}}$, which also named zero-point length, play a critical role in quantum gravity\cite{Nicolini_2022}. Experiments so far indicate it is not possible to detect experimental procedures that measure lengths with
an accuracy greater than or about $L_{0}$. This result suggests that one could think of Planck length as some kind of “zero-point length” of spacetime. In 1997\cite{Padmanabhan_1997}, T. Padmanabhan used the idea that come from T-duality(i.e., $R\leftrightarrow \alpha'/R$, $n\leftrightarrow w$) in String theory. The basic idea is he given an assumption that the path integral amplitude is invariant under the ‘duality’ transformation $ds\to L_{0}^{2}/ds$, one will modify Feynman's propagator and he showed that this propagator is the same as the one obtained by assuming that: the modification of the spacetime interval
$(x-y)^{2}$ to $(x-y^{2})+L_{0}^{2}$. One advantage of this theory is that the propagator becomes UV finite\cite{smailagic2003string}. For example, in the recent paper\cite{Gaete_2022}, they calculate the electric field strength of a point charged particle in the path integral duality and there is no singularity while the point charge particle in standard Maxwell's equations will give a singularity. By the way, a similar approach is used to obtain generalized uncertainty relationship\cite{GARAY_1995}\cite{Kempf_1995}\cite{Maggiore_1993}.\\
Usually, the approach to QFT is based on the canonical formalism,
in which a field is an operator-valued distribution. In this
approach, the derivation of the generating functional from which
we obtained the perturbative rules was arguably rather
cumbersome, even in the simple setting of a scalar field theory.
Functional quantization, also known as the path integral
formalism, is an alternative quantization procedure that
considerably simplifies the algebraic manipulations, and provides
some rather intuitive insights into what makes a theory quantum.
Quantum anomaly is the phenomenon that some symmetries of
Lagrangian are broken by quantum corrections. And under path
integral, if the transformation changes the path integral measure,
it would be anomaly.
Therefore, it is worth to quantum anomalies
under path integral.\\
In this paper, I first introduce the basic theory which has existed in previous papers by using dimensional compactification on propagators and a gauge-invariant but Lorentz-violating QED
effective action which can be obtained through the order power of the external momentum $p^{\mu}$ in Section \ref{2}.\\
In Section \ref{3}, I use the path integral to calculate the gauge invariance but Lorentz-violating QED effective action's anomalous term. 
\\
In Section \ref{4}, I give the relation between compact bulk scalar propagator and path integral duality by using some mathematical tricks, I think this relation is interesting because it gives a unification that compact bulk scalar propagate is consistent with path integral duality.

\section{BASIC THEORY}\label{2}
\subsection{Compactification of extra dimensions}
Consider a $5D$ spacetime $\mathcal{M}^{1,3}\times S^{1}$. To describe this we begin from $5D$ Minkowski space $\mathcal{M}^{1,4}$ with coordinates
\begin{equation}
x^{M}=\left(x^{\mu}, x^{4}\right),\quad M=0,...,4. \quad \mu=0,...,3.
\end{equation}
-with the metric $\eta_{MN}=$diag$(+,-,-,-,-)$ and obtain an $S^{1}$ by periodically identifying the $x^{4}$ coordinate, $x^{4}\sim  x^{4} + 2\pi R$. It’s convenient to describe this identification as
\begin{equation}
x^{M}\sim x^{M}+A^{M},\quad A^{M}=(0,0,0,0,2\pi R)
\end{equation}
This coordinate define the frame for the compactification, with
an exact $SO(1,3)$ symmetry that acts on the coordinates $x^{\mu}$.
Now to describe the braneworld moves in the $x^{4}$ direction or rotates the $x^{4}$ direction we  need to transform to a new frame
with lower-case coordinates $x^{M}$ through
\begin{equation}
x^{M}=L^{M}_{N}x^{N},
\end{equation}
where $L^{M}_{N}\in SO(1,4)$. Boosted/Rotated in
the $x^{M}$ coordinates can be written via
\begin{equation}
x^{M}\approx x^{M}+a^{M},\quad a^{M}=L^{M}_{N}A^{N}.
\end{equation}
Decompose $a^{M}$ into components tangent
and normal to the brane,
\begin{equation}
a^{M}=(a^{\mu},2\pi r),
\end{equation}
where $2\pi r$ is a fifth component scalar on the brane and $r$ is related to $R$ but also depends on the motion. Some form of $r$ in timelike/spacelike/lightlike situation has been given in \cite{kabat2023induced}. And usually we use the new defined quantity $b^{\mu}=\frac{a^{\mu}}{2\pi r}$.
\subsection{compact bulk scalar propagator}
In a general $(d+1)$-dimensional space, coordinates can be splitted as $x^{n}=(x, \xi)$ where $\xi$ is the compact dimension, $n=0,1,...,d$ and split the momenta as $k^{n}=(k, q)$ where $q$ is the momentum on $\xi$ direction. Consider a bulk scalar field of mass $m$ and denote the retarded Green’s function $G^{(d+1)}_{R}(x, \xi)$ where $d+1$ is the number of space-time dimensions and $R$ is the radius of the circle. For the standard scalar field, the Green's function is
 \begin{equation}
(\square-\mu^{2})G^{(d+1)}_{\infty}(x, \xi)=\delta^{d}(x)\delta(\xi),
 \end{equation}
where $\mu$ is the mass of a bulk scalar, this Green's function can be represented in the form
\begin{equation}
G^{(d+1)}_{\infty}(x, \xi)=\int \frac{d^{d}k}{(2\pi)^{d}}\frac{dq}{2\pi} \frac{ie^{-ik\cdot x+iq\xi}}{k^{2}-q^{2}-\mu^{2}+i\epsilon},
\end{equation}
A sum over winding numbers to compactify the $\xi$ direction,
\begin{equation}
G^{(d+1)}_{R}(x, \xi)=\sum_{w\in \mathbb{Z}}G^{(d+1)}_{\infty}(x, \xi-2\pi Rw).
\end{equation}
If the contribution such that the winding sum becomes continuous(i.e., $\sum_{w\in \mathbb{Z}}\to \int d\omega$), then the winding sum will lead to
\begin{equation}
\int dw e^{-iq2\pi Rw}=\frac{1}{R}\delta(q),
\end{equation}
which means
\begin{equation}\label{A8}
G^{(d+1)}_{R}\to \frac{1}{2\pi R}G^{(d)}_{\infty}=\frac{1}{2\pi R}\int\frac{d^{d}k}{(2\pi)^{d}} \frac{ie^{-ik\cdot x}}{k^{2}-\mu^{2}+i\epsilon}
\end{equation}
This equation nicely illustrate the relation between the compactifiaction of $(d+1)$-dimension and noncompact $d$-dimension.\\
Let's go back to focus on bulk propagation and it's convenient to set $\xi=0$ and we only consider the winding number integral(i.e., ignore $\int\frac{d^{d}k}{(2\pi)^{d}}e^{-ik\cdot x}$)
\begin{eqnarray}
\Delta=\sum_{w=-\infty}^{\infty}\int\frac{dq}{2\pi} \frac{i}{k^{2}-q^{2}-\mu^{2}+i\epsilon}e^{-iq2\pi Rw}.
\end{eqnarray}
Switching from a sum over windings to a sum over Kaluza-Klein momentum using the Poisson resummation identity\cite{kabat2023induced}
\begin{equation}
\sum_{w=-\infty}^{\infty}\int\frac{dq}{2\pi} f(q)e^{-iq2\pi Rw}=\frac{1}{2\pi R}\sum_{w=-\infty}^{\infty}f(\frac{n}{R}),
\end{equation}
this identity can put \eqref{A8} into another form
\begin{equation}\label{A10}
\Delta=\frac{1}{2\pi R}\sum_{n=-\infty}^{\infty} \frac{i}{k^{2}-(n/R)^{2}-\mu^{2}+i\epsilon}.
\end{equation}
Now let's meet a special case: set $d=4$ and consider $(x^{\mu},\xi)\approx (x^{\mu}+a^{\mu},\xi+2\pi r)$.
The bulk propagator in this case will be in the form
\begin{equation}\label{A14}
\Delta=\frac{1}{2\pi r}\sum_{n=-\infty}^{\infty} \frac{i}{k^{2}-(k\cdot b+\frac{n}{r})^{2}-\mu^{2}+i\epsilon}.
\end{equation}
$b^{\mu}\neq 0$ will break Lorentz invariance.
\subsection{A gauge-invariant but Lorentz-violating QED effective action}
Consider the action
\begin{eqnarray}
S=&&\int d^{5}x\left[\frac{1}{2}\partial_{M}\chi\partial^{M}\chi-\frac{1}{2}\mu^{2}\chi^{2}\right]\nonumber\\&&+\int d^{4}x\left[\overline{\psi}(i\slashed{\partial}-m)\psi-\lambda\overline{\psi}\psi\chi\big|_{\xi=0}\right].
\end{eqnarray}
This action means there is a real bulk scalar field $\xi$ of mass $\mu$ that has a
Yukawa coupling to the electron and $\lambda$ is a coupling constant.\\
One-loop electron self-energy arising from a Yukawa coupling to a bulk scalar and it's bulk propagator is\cite{kabat2023induced}
\begin{eqnarray}
i\sum=&&\frac{\lambda^{2}}{2\pi r}\sum_{n=-\infty}^{\infty}\int\frac{d^{4}k}{(2\pi)^{4}}\frac{\slashed{k}+m}{k^{2}-m^{2}+i\epsilon}\times\nonumber\\&& \frac{1}{(k-p)^{2}-((k-p)\cdot b+\frac{n}{r})^{2}-m^{2}+i\epsilon}
\end{eqnarray}
Expand this propagator in powers of the external momentum $p$ and notice the first order in $p^{\mu}$, this term will produce a gauge-invariant but Lorentz-violating term in the effective action
\begin{equation}\label{A12}
\mathcal{L}=ic_{\mu\nu}\overline{\psi}\gamma^{\mu}D^{\nu}\psi,
\end{equation}
$D^{\nu}=\partial^{\nu}-ieA^{\nu}$ to preserve the gauge-invariant property and the coefficient $c_{\mu\nu}$ is
\begin{equation}
c_{\mu\nu}=-\frac{1}{16\pi^{2}}\frac{\lambda^{2}}{\pi r}\left(b_{\mu}b_{\nu}-\frac{1}{4}\eta_{\mu\nu}b^{2}\right)I_{1},
\end{equation}
\begin{eqnarray}
&&I_{n}=\nonumber\\&&\frac{1}{\sqrt{\pi}}\sum_{w=-\infty}^{\infty}\int_{0}^{\infty}ds\int_{0}^{\infty}dt\frac{s^{n}w^{2}}{\sqrt{t}(s+t)^{4}}\times\nonumber\\&& exp\left\{-s(\pi mr)^{2}-t(\pi\mu r)^{2}-\frac{s+t(1-b^{2})}{t(s+t)}w^{2} \right\}
\end{eqnarray}
\section{Anomalous function of a 4-dimensional Lorentz-violating QED effective action}\label{3}
\subsection{General considerations}
Before we calculate the anomaly term of the Lagrangian \eqref{A12}, let's do some preparation. Without loss of generality, consider a set of fermion fields $\psi_{n}(x)$, which we encapsulate into a multiplet denoted $\boldsymbol{\psi}(x)$. Consider now the following transformation of the fermion fields:
\begin{equation}
\boldsymbol{\psi}(x)\to U(x)\boldsymbol{\psi}(x)
\end{equation}
The Hermitic conjugate of $\boldsymbol{\psi}(x)$ transforms as $\boldsymbol{\psi}^{\dagger}(x)\to \boldsymbol{\psi}(x)U^{\dagger}(x)$, so that we have
\begin{equation}
\overline{\boldsymbol{\psi}}(x)=\boldsymbol{\psi}^{\dagger}\gamma^{0}\to \boldsymbol{\psi}^{\dagger}(x)U^{\dagger}(x)\gamma^{0}=\overline{\boldsymbol{\psi}}(x)\gamma^{0}U^{\dagger}(x)\gamma^{0}.
\end{equation}
The measure is transformed with the inverse of the determinant of the transformation,
\begin{equation}
\mathcal{D}[\boldsymbol{\psi}]\mathcal{D}[\overline{\boldsymbol{\psi}}]\to\frac{1}{det(U)det(\overline{U})}\mathcal{D}[\boldsymbol{\psi}]\mathcal{D}[\overline{\boldsymbol{\psi}}],
\end{equation}
where the matrices $U$ and $\overline{U}$ carry both indices for the fermion species and space-time indices:
\begin{equation}
U_{xm,yn}=U_{m,n}(x)\delta(x-y),
\end{equation}
\begin{equation}
\overline{U}_{xm,yn}=\left(\gamma^{0}U^{\dagger}(x)\gamma^{0}\right)_{m,n}\delta(x-y).
\end{equation}
Split the spinor into right-handed and left-handed projections,
\begin{equation}
\boldsymbol{\psi}_{R}=\left(\frac{1+\gamma^{5}}{2}\right)\boldsymbol{\psi},\quad \boldsymbol{\psi}_{L}=\left(\frac{1-\gamma^{5}}{2}\right)\boldsymbol{\psi},
\end{equation}
and consider a chiral transformation
\begin{equation}
U(x)=e^{i\alpha(x)\gamma^{5}t},
\end{equation}
where $t$ is a Hermitian matrix that does not contain Dirac matrices and $\alpha(x)$ is pesudofunction. And
\begin{equation}
\gamma^{0}U^{\dagger}(x)\gamma^{0}=\gamma^{0}e^{-i\alpha(x)\gamma^{5}t}\gamma^{0}=e^{i\alpha(x)\gamma^{5}t}=U(x).
\end{equation}
Thus, $\overline{U}=U$, and $detU=det\overline{U}$. To calculate anomaly function we need to let $detU\neq 1$ thus the measure is not invariant and transforms according to
\begin{equation}
\mathcal{D}[\boldsymbol{\psi}]\mathcal{D}[\overline{\boldsymbol{\psi}}]\to\frac{1}{(detU)^{2}}\mathcal{D}[\boldsymbol{\psi}]\mathcal{D}[\overline{\boldsymbol{\psi}}].
\end{equation}
And
\begin{equation}
\frac{1}{(detU)^{2}}=exp\left(-2\text{Tr} ln U\right)=exp\left(i\int d^{4}x\alpha(x)\mathcal{A}(x)\right),
\end{equation}
where $\mathcal{A}(x)$ is called the anomaly function and $\mathcal{A}(x)=-2\delta(x-x)tr(\gamma^{5}t)$. In terms of this function, the measure transforms as
\begin{equation}
\mathcal{D}[\boldsymbol{\psi}]\mathcal{D}[\overline{\boldsymbol{\psi}}]\to e^{i\int d^{4}x\alpha(x)\mathcal{A}(x)}\mathcal{D}[\boldsymbol{\psi}]\mathcal{D}[\overline{\boldsymbol{\psi}}].
\end{equation}
This measure can be absorbed into a redefinition of Langrangian,
\begin{equation}
\mathcal{L}(x)\to \mathcal{L}(x)+i\alpha(x)\mathcal{A}(x).
\end{equation}
\subsection{Calculation of $\mathcal{A}(x)$}
Manipulate finite expressions, we must regularize the delta function. This can be done by
\begin{equation}
\label{QA513}
\mathcal{A}(x)=-2\lim_{y\to x, M\to +\infty}\text{Tr}\left\{\gamma^{5}t \mathcal{F}\left(-\frac{(c_{\mu\nu}\gamma^{\mu}D^{\mu}_{x})^{2}}{M^{2}}\right) \right\} \delta(x-y),
\end{equation}
where $D_{x\mu}=\partial_{\mu}-ieA_{\mu}(x)$ and $\mathcal{F}(s)$ is called regulator and satisfies
\begin{equation}
\mathcal{F}(0)=1,\quad \mathcal{F}(+\infty)=0,\quad s\mathcal{F}'(s)=0 \quad at\quad s=0,+\infty.
\end{equation}
Then, we replace the delta function by its Fourier representation,
\begin{equation}
\delta(x-y)=\int\frac{d^{4}k}{(2\pi)^{4}}e^{ik\cdot(x-y)},
\end{equation}
which leads to
\begin{eqnarray}
\mathcal{A}(x)&&=-2\int\frac{d^{4}k}{(2\pi)^{4}}\lim_{M\to +\infty}\nonumber\\&&\times \text{Tr}\left\{\gamma^{5}t \mathcal{F}\left(-\frac{(ic_{\mu\nu}\gamma^{\mu}k^{\nu}+c_{\mu\nu}\gamma^{\mu}D^{\mu}_{x})^{2}}{M^{2}}\right) \right\}
\end{eqnarray}
where we have used the identity\cite{F19}
\begin{equation}
\lim_{y\to x}\mathcal{F}(\partial_{x})e^{ik(x-y)}=\mathcal{F}(ik+\partial_{x}).
\end{equation}
Redefining the integration variable, $k\to Mk$:
\begin{eqnarray}
&&\mathcal{A}(x)=-2M^{4}\int\frac{d^{4}k}{(2\pi)^{4}}\lim_{M\to +\infty}\times\text{Tr}\left\{\gamma^{5}t \mathcal{F}\left(B\right) \right\},    
\end{eqnarray}
\begin{equation}
B=c_{\mu\nu}c_{\sigma}^{\mu}k^{\nu}k^{\sigma}-2i\frac{c_{\mu\nu}c_{\sigma}^{\mu}k^{\nu}D_{x}^{\sigma}}{M}-\frac{(c_{\mu\nu}\gamma^{\mu}D_{x}^{\nu})^{2}}{M^{2}}
\end{equation}
where we have used the anticommutation relation $\{\gamma^{\mu},\gamma^{\nu}\}=2g^{\mu\nu}$.\\ Expand the function $\mathcal{F}(\cdot)$ in powers of $1/M$. Let's consider the non-$M^{4}$ term
\begin{equation}
\mathcal{A}(x)=-2\int\frac{d^{4}k}{(2\pi)^{4}}\mathcal{F}''(c_{\mu\nu}c_{\sigma}^{\mu}k^{\nu}k^{\sigma})\text{Tr}\left\{\gamma^{5}t \left(c_{\mu\nu}\gamma^{\mu}D_{x}^{\nu}\right)^{4} \right\}.
\end{equation}
Define $l_{\nu}=c_{\mu\nu}k^{\mu}$ and consider $det(c_{\mu\nu})$ is a constant, Then by Wick rotation $l^{0}=i\kappa$, we obtain
\begin{equation}
\int d^{4}k \mathcal{F}''(c_{\mu\nu}c_{\sigma}^{\mu}k^{\nu}k^{\sigma})=\int d^{4}l \frac{1}{det(c_{\mu\nu})}\mathcal{F}''(l^{2})=\frac{i\pi^{2}}{det(c_{\mu\nu})},
\end{equation}
and $\left(c_{\mu\nu}\gamma^{\mu}D_{x}^{\nu}\right)^{2}=c_{\mu\nu}c^{\mu}_{\sigma}D^{\nu}D^{\sigma}-\frac{ie}{4}c_{\mu\nu}c_{\rho\sigma}F^{\mu\rho}\left[\gamma^{\nu},\gamma^{\sigma}\right]$. Using the identity
\begin{equation}
\text{Tr}\left(\gamma^{5}\gamma_{\mu}\gamma_{\nu}\gamma_{\rho}\gamma_{\sigma}\right)=4i\epsilon_{\mu\nu\rho\sigma},
\end{equation}
we obtain
\begin{equation}
\mathcal{A}(x)=-\frac{e^{2}}{16\pi^{2}det(c)}\epsilon^{\mu\rho\alpha\gamma}\text{Tr}\left(tc_{\mu\nu}c_{\rho\sigma}c_{\alpha\beta}c_{\gamma\lambda}F^{\nu\sigma}F^{\beta\lambda}\right)
\end{equation}
Here matrix $t$ can be act on flavors. For example, we can consider u and d quarks sector. But here we just consider one of these cases:
\begin{equation}
\mathcal{A}(x)=-\frac{e^{2}}{16\pi^{2}det(c)}\epsilon^{\mu\rho\alpha\gamma}c_{\mu\nu}c_{\rho\sigma}c_{\alpha\beta}c_{\gamma\lambda}F^{\nu\sigma}F^{\beta\lambda}.
\end{equation}
\section{Relation between compact bulk scalar propagator and Path Integral duality}\label{4}
Let's go back to \eqref{A10} and recover the Green function's form and for the convenience of the following discussion, we may wish to change the metric notation to $\eta^{new}_{MN}=$diag$(-,+,+,+,+)$, the change of metric will have an effect $-(n/R)^{2}\to (n/R)^{2}$ and $-\mu^{2}\to \mu^{2}$
\begin{equation}
G^{(d+1)}_{R}=\frac{i}{2\pi R}\sum_{n=-\infty}^{\infty}\int\frac{d^{d}k}{(2\pi)^{d}}\frac{e^{-ik\cdot x}}{k^{2}+(n/R)^{2}+\mu^{2}}.
\end{equation}
Let's deal this equation by constituting the Schwinger’s proper time version of the propagator,
\begin{eqnarray}
G^{(d+1)}_{R}=\frac{i}{2\pi R}\sum_{n=-\infty}^{\infty}\int&&\frac{d^{d}k}{(2\pi)^{d}}e^{-ik\cdot x}\int^{\infty}_{0}ds\nonumber\\&&\times e^{-s\left(k^{2}+(n/R)^{2}+\mu^{2}\right)}.
\end{eqnarray}
The integral of $k^{\mu}$ is a Gaussian quadratic integral, so 
\begin{equation}
G^{(d+1)}_{R}=\frac{i}{2\pi R}\sum_{n=-\infty}^{\infty}\int^{\infty}_{0}\frac{ds}{(2\pi)^{d/2}}exp\left\{-\frac{x^{2}}{s}-s\left[(n/R)^{2}+\mu^{2}\right]\right\}.   
\end{equation}
For a clearer view, it is advisable to discretize the integral of $s$: $\int_{0}^{\infty}ds\to \sum_{s\in \mathbb{N}}$
\begin{equation}\label{IV1}
G^{(d+1)}_{R}=\frac{i}{2\pi R}\sum_{n\in\mathbb{Z},s\in\mathbb{N}}\frac{1}{(2\pi)^{d/2}}exp\left\{-\frac{x^{2}}{s}-s\left[(n/R)^{2}+\mu^{2}\right]\right\}. 
\end{equation}
when $s=1$, $n=1$ and we consider $\frac{1}{R}=L_{0}$ where $L_{0}$ is the zero-point length, \eqref{IV1} can be explained as we add a zero point length to any spacetime interval(i.e., modification of the spacetime interval $(x-y)^{2}\to (x-y)^{2}+L_{0}^{2}$), so we actually get the same result with the result in path integral duality.

\section{conclusion}
This work mainly completed two things: (i) calculate the anomalous function of $\mathcal{L}=i\overline{\psi}c_{\mu\nu}\gamma^{\mu}D^{\nu}\psi$ through path integration, and (ii) discover that the compact bulk scalar propagator and path integral duality are consistent. The second discover is quite interesting, as this relation directly brings new perspective to future research: the behavior of breaking Lorentz invariance caused by dimensional compactness can be seen as path integral duality.

\begin{acknowledgments}
Thanks \textbf{Dan Wohns in Perimeter Institute}. In Perimeter Scholars International(PSI) START Program's mini-project 2023, he provides me the project: Quantum anomalous effects under path integral. Therefore, I started to know how to use path integral to calculate the anomaly.\\
Thanks \textbf{Tigao Adorno in Xi'an Jiaotong-Liverpool University} to support help and checked $\alpha(x)$ needs to be pseudo to preserve the parity invariant symmetry of
the action.\\
\end{acknowledgments}

\nocite{*}

% Produces the bibliography via BibTeX.
\providecommand{\noopsort}[1]{}\providecommand{\singleletter}[1]{#1}%
\end{document}